\documentclass[reprint,prl,aps,twocolumn,floatfix,superscriptaddress]{revtex4-1}
\usepackage{eurosym}
\usepackage{amsmath,amsfonts,amssymb,graphics,graphicx,color,bm}
\usepackage{hyperref}
\usepackage[percent]{overpic}
\usepackage{braket}
\usepackage{accents}
\usepackage{tikz}

\begin{document}

\title{Translation-Ivariant Parent Hamiltonians of Valence Bond Crystals}

\author{Daniel Huerga}
\email{huerga@itp3.uni-stuttgart.de}
\affiliation{Institut f\"ur Theoretische Physik III, Universit\"at Stuttgart, Pfaffenwaldring 57, D-70550 Stuttgart, Germany}
\author{Andr\'es Greco}
\author{Claudio Gazza}
\affiliation{Instituto de F\'isica Rosario, CONICET, and Facultad de Ciencias Exactas Ingenier\'ia y Agrimensura, Universidad Nacional de Rosario, Boulevard 27 de Febrero 210 bis, 2000 Rosario, Argentina}
\author{Alejandro Muramatsu}
\affiliation{Institut f\"ur Theoretische Physik III, Universit\"at Stuttgart, Pfaffenwaldring 57, D-70550 Stuttgart, Germany}

\begin{abstract}
We present a general method to construct translation-invariant and SU(2) symmetric antiferromagnetic parent Hamiltonians of valence bond crystals (VBCs).
The method is based on a canonical mapping transforming $S=1/2$ spin operators into a bilinear form of a new set of {\it dimer fermion} operators.
We construct parent Hamltonians of the columnar- and the staggered-VBC on the square lattice, for which the VBC is an eigenstate in all regimes and the exact ground state in some region of the phase diagram.
We study the depart from the exact VBC regime upon tuning the anisotropy by means of the hierarchical mean field theory and exact diagonalization on finite clusters.
In both Hamiltonians, the VBC phase extends over the exact regime and transits to a columnar antiferromagnet (CAFM) through a window of intermediate phases, revealing an intriguing competition of correlation lengths at the VBC-CAFM transition.
The method can be readily applied to construct other VBC parent Hamiltonians in different lattices and dimensions.
\end{abstract}

\maketitle
Quantum magnets host a wealth of phases and exotic phenomena.
The complex interplay between spin interactions and lattice topology may eventually prevent the stabilization of magnetic order.
In particular, frustrated antiferromagnetic spin-1/2 interactions may favor the partition of the system into nearest-neighbor (NN) spin singlets, so-called {\it valence-bonds} (VB), covering the lattice in a periodic pattern or VB crystal (VBC)~\cite{Misguich2012}.
Eventually, VBs may resonate and recover translational invariance by forming a resonating-VB (RVB) {\it spin liquid}, as has been conjectured to occur in high-T$_c$ cuprates~\cite{Fazekas1974,Anderson1987}.
Particularly interesting is the zero-temperature quantum phase transition from the former state to an ordered AF phase~\cite{Sachdev2008}, a transition that can be experimentally probed upon tuning external pressure~\cite{Ruegg2008} or magnetic field on various materials~\cite{Zapf2014}.
Under certain specific conditions, it has been argued that a class of VBC-AF transitions in two-dimensions (2D) are driven by the deconfinement of fractional excitations at a critical point~\cite{Senthil2004}, contrary to what is generally expected within Landau theory.
To test these ideas, a family of antiferromagnetic Heisenberg Hamiltonians with additional four- and six-spin interactions favoring VBC order and amenable to quantum Monte Carlo (QMC) computations has been introduced during the last decade~\cite{Sandvik2007,Lou2009_AF,Pujari2013}. 
These models, so-called $J-Q$ models, show unusual scaling behavior at the VBC-AF transition point~\cite{Jiang2008,Sandvik2010,Pujari2013,Chen2013,Shao2016} and, in some cases, a weak VBC signal~\cite{Sandvik2007,Jiang2008}.
In order to furnish this putative new class, it would be desirable to enlarge the number of Hamiltonians hosting VBC-AF transitions.
In particular, antiferromagnetic Hamiltonians hosting an {\it exact} VBC ground-state (GS) shall provide a convenient test bed where the VBC phase is unambiguously defined.

There exist few parent Hamiltonians of VBC states. 
The paradigmatic models in 1D and 2D those of Majumdar-Gosh~\cite{Majumdar1969} and Shastry-Sutherland~\cite{Shastry1981}, respectively. 
More recently, 2D parent Hamiltonians of different VBCs have been constructed~\cite{Kumar2005,Gelle2008} by using sums of local projectors, so-called Klein models~\cite{Klein1982}. 
However, this method may lead to a GS manifold of various degenerate VBC patterns~\cite{Batista2004}, a feature that has been otherwise exploited to construct parent Hamiltonians of short-range RVBs~\cite{Fujimoto2005,Seidel2009,Cano2010}, important for their potential applications to topological quantum computation~\cite{Kitaev2003}.

\begin{figure}[t]
\includegraphics[width=0.48\textwidth]{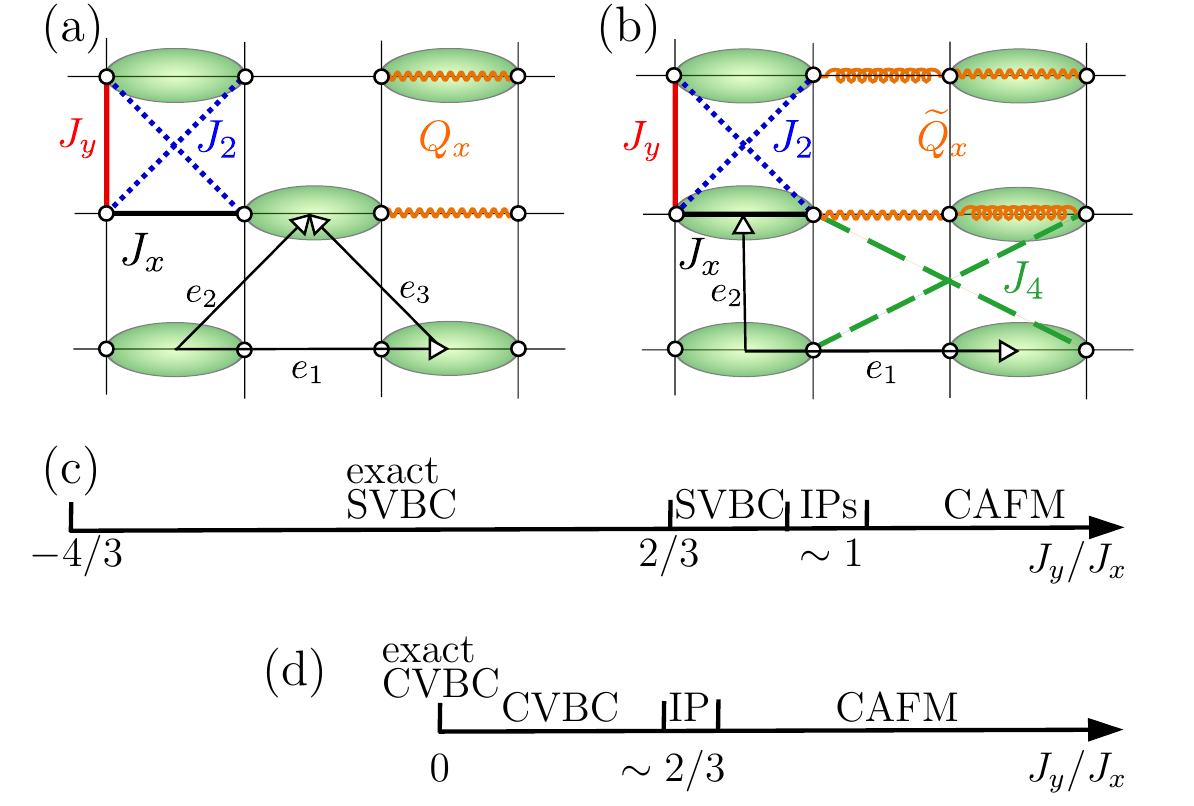}
\caption{Schematic representation of the SVBC (a) and CVBC (b) states and their respective parent Hamiltonians, together with their phase diagrams, (c) and (d). Green ovals represent dimers, thick solid lines represent the Heisenberg interaction among NN spins along the $x$ (black) and $y$ (red) directions, second-NN spins (dotted blue), and a subset of fourth-NN spins (dashed green). Pairs of orange waved- and spring-lines represent anisotropic four-spin interactions.\label{fig:intro}}  
\end{figure}

In this Letter, we present a general method to construct SU(2) symmetric and translation-invariant VBC parent Hamiltonians based on a canonical mapping that exactly identifies a VBC state with the vacuum of a new set of \textit{dimer fermions} (DF). 
Starting with a generic anisotropic AF Heisenberg Hamiltonian as an Ansatz, this mapping guides our search for additional translation-invariant interactions that, summed to the Ansatz, make the vacuum an exact eigenstate of the sum.
As an example, we construct parent Hamiltonians of the staggered- and the columnar-VBC (SVBC and CVBC) on the square lattice, finding that anisotropic four-spin interactions are required to exactly stabilize the VBC in both cases.
Interestingly, the SVBC parent Hamiltonian hosts an exact GS within a finite region of the phase diagram, differently from other parent Hamiltonians where the exact GS is found on a single point~\cite{Majumdar1969,Batista2004,Kumar2005,Gelle2008}. 
Upon tuning the anisotropy, the VBC transits to a columnar AFM (CAFM) phase characterized by a finite magnetization at the columnar wave vector $(0,\pi)$ through a window of intermediate phases (IPs) characterized by the strong competition of correlations at different characteristic lengths (see Fig.~\ref{fig:intro}).

\textit{Dimer fermion mapping}.--- Consider a lattice where a quantum spin $S=1/2$, described by the operator $S^a_j$ ($a=x,y,z$), resides on each vertex, $j$, of a lattice of size $\mathcal N$. 
We choose a VBC pattern on this lattice, defining a superlattice where in each vertex, $r$, resides a dimer, i.e. a pair of NN spins.
Each site of the lattice can thus be referred to by its position within a dimer, $\alpha$, and the position of the dimer in the superlattice, i.e. $j\equiv r,\alpha$ ($\alpha=1,2$). 
Considering the Majorana representation of spins \cite{Tsvelik1992}, 
$
S^a_j=-\frac{\rm i}{2}\sum_{b,c} \varepsilon^{abc}\eta^b_j\eta^c_j,
$
where $\varepsilon^{abc}$ is the Levi-Civita tensor and the Majorana fermions, $\eta_j^a=\eta^{a\dag}_j$, obey anticommutation relations $\lbrace \eta^a_i,\eta^b_j\rbrace=\delta^{ab}\delta_{ij}$,
we define a \textit{dimer fermion} (DF) as the linear combination of two Majorana fermions of a dimer,
\begin{equation}
f^{a\dag}_{r}= \frac{1}{\sqrt{ 2}}\left(\eta_{r,1}^a + {\rm i}\eta_{r,2}^a\right),~~~f^a_{r}=(f^{a\dag}_{r})^\dag\label{eq:fermion_def}.
\end{equation}
The DFs obey the usual fermionic anticommutation relations, $\lbrace f^a_{r},f^{b\dag}_{r'}\rbrace=\delta^{ab}\delta_{rr'}$. Inverting the relation (\ref{eq:fermion_def}), we can express the spin operators in terms of DFs,
\begin{eqnarray}
S^a_{r,1}&=&\frac{1}{2}\sum_{b,c}\varepsilon^{abc}\left(T^{bc}_{r}+\Delta^{bc}_{r}\right),\label{eq:sf1}\\
S^a_{r,2}&=&\frac{1}{2}\sum_{b,c}\varepsilon^{abc}\left(T^{bc}_{r}-\Delta^{bc}_{r}\right),\label{eq:sf2}
\end{eqnarray}
where we have defined the fully antisymmetric number-conserving and non-conserving DF operators,
\begin{eqnarray}
T^{ab}_{r}&=&-\frac{\rm i}{2}\left( f^{a\dag}_{r}f^b_{r}-f^{b\dag}_{r}f^a_{r} \right)\label{eq:T},\\
\Delta^{ab}_{r}&=&-\frac{\rm i}{2}\left( f^{a\dag}_{r}f^{b\dag}_{r}-f^b_{r}f^a_{r} \right)\label{eq:D}.
\end{eqnarray}
The operators (\ref{eq:T}) and (\ref{eq:D}) are indeed the generators of the SU(2)$\otimes$SU(2)$\simeq$SO(4) algebra of a dimer~\cite{Foussats2011}. Notice that the mapping (\ref{eq:sf1})-(\ref{eq:sf2}) relating S=1/2 spin operators and DFs is canonical without the need of any additional constraint, contrary to other slave-particle mappings.

In the DF representation, the dimer Hilbert space is doubled with respect the physical one, existing two equivalent copies of the singlet $(\ket{s})$ and the triplet states. 
By applying the total spin operator, $\mathbf{S}_r^2=(\mathbf{S}_{r,1}+\mathbf{S}_{r,2})^2$, 
and its third component, $S_r^z=(S^z_{r,1}+S^z_{r,2})$, to the eight DF states of the dimer we can identify the vacuum and double occupied DF states, on the one hand, and the fully and single occupied states, on the other, with two equivalent copies of the singlet and triplet states (see Supplemental Material~\footnote{see Supplemental Material for details on the dimer fermion mapping and on the HMFT and ED computations} for details). Without loss of generality, the VBC state can be identified with the DF vacuum,
\begin{equation}
\ket{\text{VBC}}=\prod_r \ket{s}_r=\ket{0}\label{eq:vbs}.
\end{equation}

A direct consequence of the DF mapping (\ref{eq:sf1})-(\ref{eq:sf2}) is that the Heisenberg (HB) interaction among two spins, $B_{ij}=\mathbf{S}_i\mathbf{S}_j$, maps to a one-body DF operator,
\begin{equation}
D_{r}= -\frac{3}{4} + \frac{1}{2}n_r(3-n_r),\label{eq:intra}
\end{equation}
where $n_r= \sum_a f^{a\dag}_rf^{a}_r$, when it acts on the two spins comprising a dimer ($i=r,1;~j=r,2$). Alternatively, $B_{ij}$ maps to a two-body DF operator when it acts on two different dimers ($i\in r,~j\in r'\ne r$),
\begin{eqnarray}
C^{\alpha\alpha'}_{rr'}&=&\frac{1}{2}\sum_{a,b} (
T_r^{ab}T_{r'}^{ab}
+ V^{\alpha\alpha'}T_r^{ab}\Delta_{r'}^{ab}\notag\\
&&~~~~~~+ V^{\alpha'\alpha}\Delta_r^{ab}T_{r'}^{ab}
+ W^{\alpha\alpha'}\Delta_r^{ab}\Delta_{r'}^{ab}),\label{eq:inter}
\end{eqnarray}
where $V^{\alpha\alpha'}$ and $W^{\alpha\alpha'}$ refer to the elements of the 2$\times$2 matrices,
\begin{equation*}
V= 
\left(
\begin{array}{cc}
1 & -1\\
1 & -1
\end{array}
\right)
~~~\text{and}~~~W=
\left(
\begin{array}{cc}
~~1 &-1\\
-1 & ~~1
\end{array}
\right),\label{eq:VW}
\end{equation*}
encoding the information about the position of the two spins within their respective dimers.

\textit{VBC parent Hamiltonian construction}.--- Once a VBC pattern is chosen in the lattice of interest, we propose a general SU(2) symmetric AFM ansatz Hamiltonian, $H_0$. For the sake of simplicity, let us restrict ourselves to the construction of the SVBC on the square lattice, although the same steps can be followed in other cases. In this case,
\begin{equation}
H_0= \sum_{j}  \left(J_x B_{j,j+\hat{x}}+J_yB_{j,j+\hat{y}}\right).\label{eq:Hansatz}
\end{equation}
where we refer by $\hat{x}$ and $\hat{y}$ to the unit vectors defining the square lattice. We rewrite (\ref{eq:Hansatz}) in terms of DFs by directly applying the DF mapping (\ref{eq:sf1})-(\ref{eq:sf2}),
\begin{eqnarray}
H_0^{\text{DF}}&=& J_x\sum_r \left(D_r+C^{21}_{r,r+e_1}  \right)\notag\\
&&+ J_y\sum_r \left(C^{21}_{r,r+e_2}
+ C^{12}_{r,r+e_3}\right)
,\label{eq:H0_DF}
\end{eqnarray}
where $e_1,e_2$ and $e_3$ refer to the basis vectors of the triangular superlattice defined by the SVBC (Fig.\ref{fig:intro}).

The terms preventing the DF vacuum to be an eigenstate of (\ref{eq:H0_DF}) are 
the $\Delta\Delta$ terms of the inter-dimer operators (\ref{eq:inter}), as $C_{rr'}^{\alpha\alpha'}\ket{0}= W^{\alpha\alpha'}\sum_{a,b}\Delta_r^{ab}\Delta_{r'}^{ab}\ket{0}\label{eq:not_eigen}.$
Therefore, we need to search for {\it additional} translation-invariant and SU(2) symmetric terms ($H_{\text{ad}}$) such that they annihilate the $\Delta\Delta$ terms in (\ref{eq:H0_DF}), and make the DF vacuum an eigenstate of the total Hamiltonian, $H=H_0+H_\text{ad}$, with eigenvalue $\varepsilon=-3J_x{\mathcal N}/8$. In constructing $H_\text{ad}$, we make use of the symmetry properties of the inter-dimer operator~(\ref{eq:inter}),
\begin{equation} C^{11}_{rr'}\ket{0}=C^{22}_{rr'}\ket{0}=-C^{12}_{rr'}\ket{0}=-C^{21}_{rr'}\ket{0}\label{eq:B11}.
\end{equation}

In particular, by adding a second-NN Heisenberg term, $J_2\sum_{\langle\langle ij\rangle\rangle}B_{ij}$, with $J_2=J_y/2$, we can annihilate the $\Delta\Delta$ terms along the $e_2$ and $e_3$ directions of the triangular superlattice (see Fig.~\ref{fig:intro}). The remaining $\Delta\Delta$ terms along the $e_1$ direction can be annihilated by adding an anisotropic four-spin interaction, $Q_x\sum_{\langle ijkl\rangle}B_{ij}B_{kl}$, with $Q_x=2J_x/3$, where $\langle ijkl\rangle$ refers to a plaquette of the square lattice. Finally, the SVBC is then an eigenstate of
\begin{eqnarray}
H&=&
\sum_{j}  \left(J_x B_{j,j+\hat{x}}+J_yB_{j,j+\hat{y}}\right)\notag\\
&&+\frac{J_y}{2}\sum_{\langle\langle ij\rangle\rangle}B_{ij}
+\frac{2J_x}{3}\sum_{\langle ijkl\rangle}B_{ij}B_{kl}\label{eq:Hstag}.
\end{eqnarray}

In order to assess whether the DF vacuum is not just an eigenstate but the exact GS in any regime, the total Hamiltonian is expressed as a translation-invariant sum of local Hamiltonians shifted by the DF vacuum eigenvalue, i.e. $H=\varepsilon+\sum_j H_j$. The local Hamiltonian of (\ref{eq:Hstag}) is straightforward and thus given in the Supplemental Material. If the local Hamiltonian $H_j$ is semi-positive definite, the DF vacuum is the exact GS. In particular, the SVBC is the exact GS of (\ref{eq:Hstag}) within the range $-4/3\le J_y/J_x \le 2/3$. 

Notice that, alternatively to the four-spin interaction, we may have added a second-NN HB interaction along the $x$-direction to annihilate the $\Delta\Delta$ terms along the $e_1$ direction. However, the resulting local Hamiltonian is not semi-definite positive.

Following the same steps, one can show that the CVBC is the exact GS of 
\begin{eqnarray}
H&=&
\sum_{j}  \left(J_x B_{j,j+\hat{x}}+J_yB_{j,j+\hat{y}}\right)
+J_2\sum_{\langle\langle ij\rangle\rangle}B_{ij}
\notag\\
&&+J_{4}\sum_{\langle\langle\langle ij\rangle\rangle\rangle}B_{ij} 
+\widetilde{Q}_x \sum_{\langle ijkl\rangle} B_{ij} B_{kl},\label{eq:Hcol}
\end{eqnarray}
for the particular point $J_y=0$, when the couplings are fixed to $J_2=J_y$, $J_4=J_y/2$, and $\widetilde{Q}_x=J_x/3$, and where the third term accounts for HB interactions among fourth-NN spins, and the fourth term accounts for four-spin interactions in tilted plaquettes (see Fig.~\ref{fig:intro}). 

\textit{Phase diagram}.--- As the four-spin AF interactions present in (\ref{eq:Hstag}) and (\ref{eq:Hcol}) pose sign-problems to state-of-the-art QMC simulations~\cite{Kaul2013}, we study the depart from the exact VBC GS by combining exact diagonalization (ED) on finite clusters with periodic boundary conditions and hierarchical mean field theory (HMFT) in the Gutzwiller approximation~\cite{Isaev2009,Huerga2014,Huerga2016}. 
From the technical standpoint, both methods involve the diagonalization of finite clusters of $N$ sites, and their combined use provide complementary information about the thermodynamic limit.

The HMFT-Gutzwiller consists in using an homogeneus product of cluster states as an Ansatz for the GS in the thermodynamic limit. 
Its variational determination reduces to perform ED on a cluster with open boundary conditions and a set of self-consistently defined mean-fields acting on its boundaries that allow for the breakdown of symmetries and stabilization of different long-range orders.
As a consequence, the sizes attained with HMFT are smaller than those of ED.
For appropriate choices of the cluster shape ---romboid and diamond for the SVBC, and square for the CVBC--- the HMFT wave function contains the exact VBC state (see Supplemental Material for details).
In addition, it allows for the systematic computation of observables directly in the thermodynamic limit. 
In particular, the energy is an upper bound to the exact one. 
We characterize VBC and CAFM phases by computing the magnetization
$
M_\mathbf{k}= \frac{1}{N}\sum_j e^{{\rm i}\mathbf{kr}_j} \langle S^z_j \rangle\label{eq:OPm},
$
at different wave vectors and the dimerization along the $x$ and $y$ directions,
$
\mathcal{D}_\nu= \frac{1}{N}\sum_j(-1)^{j_\nu} \langle S_j S_{j+\hat{\nu}}\rangle, \label{eq:OPD}
$
where $\nu=x,y$~\cite{Sachdev2008}.

\begin{figure}[t!]
\includegraphics[width=0.46\textwidth]{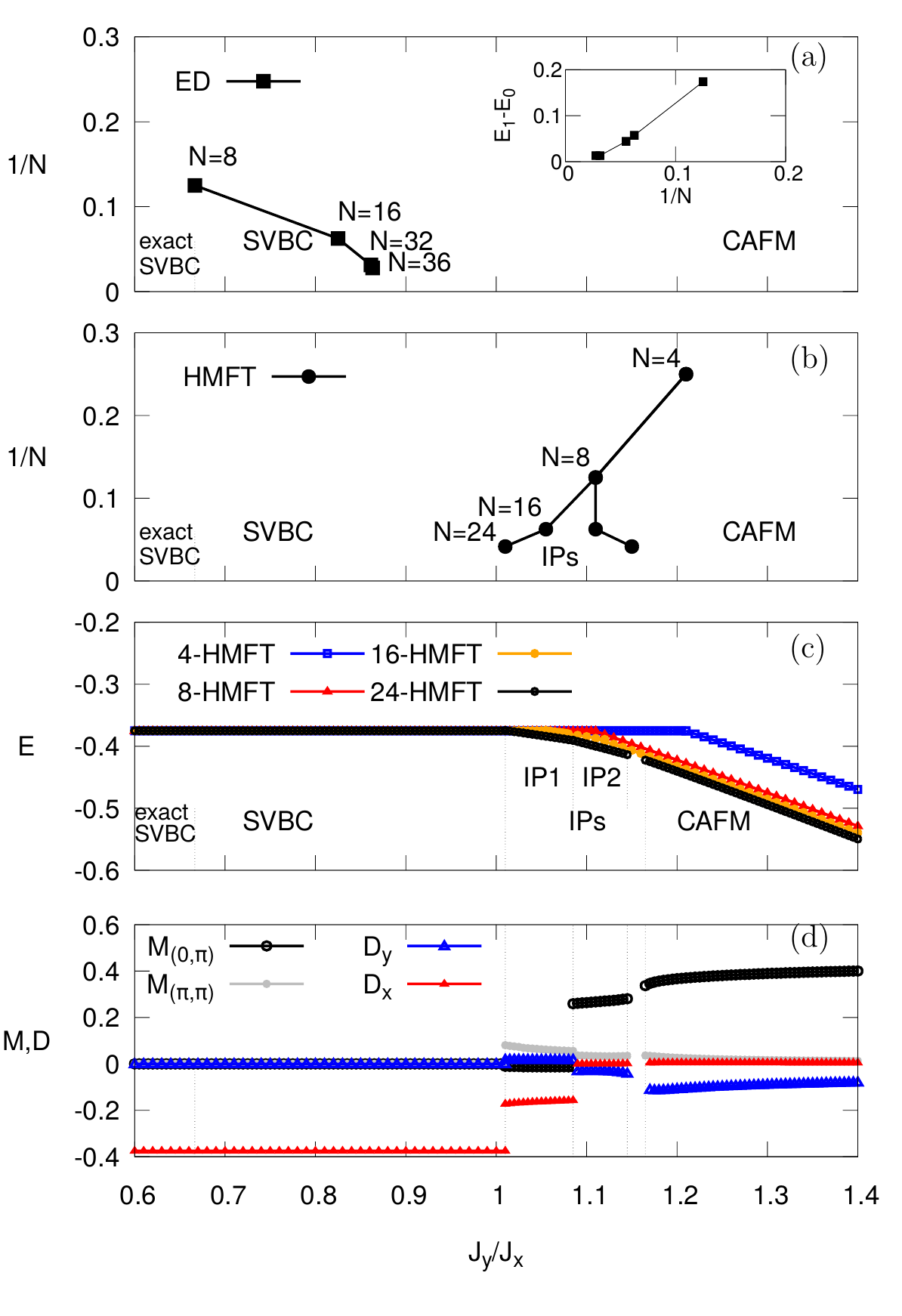}
\caption{\label{fig:DEMstag}Scaling of the GS phase diagram  of the SVBC parent Hamiltonian (\ref{eq:Hstag}) as computed with (a) ED and (b) HMFT. Inset (a): Singlet-triplet gap computed with ED at $J_y/J_x=1.4$.  (c) Energy per site (units of $J_x$) computed with HMFT for different cluster sizes of romboid shape. (d) Magnetization $M$ and dimerization $\mathcal{D}$ as computed with 24-HMFT.}
\end{figure}

\begin{figure}[t!]
\includegraphics[width=0.46\textwidth]{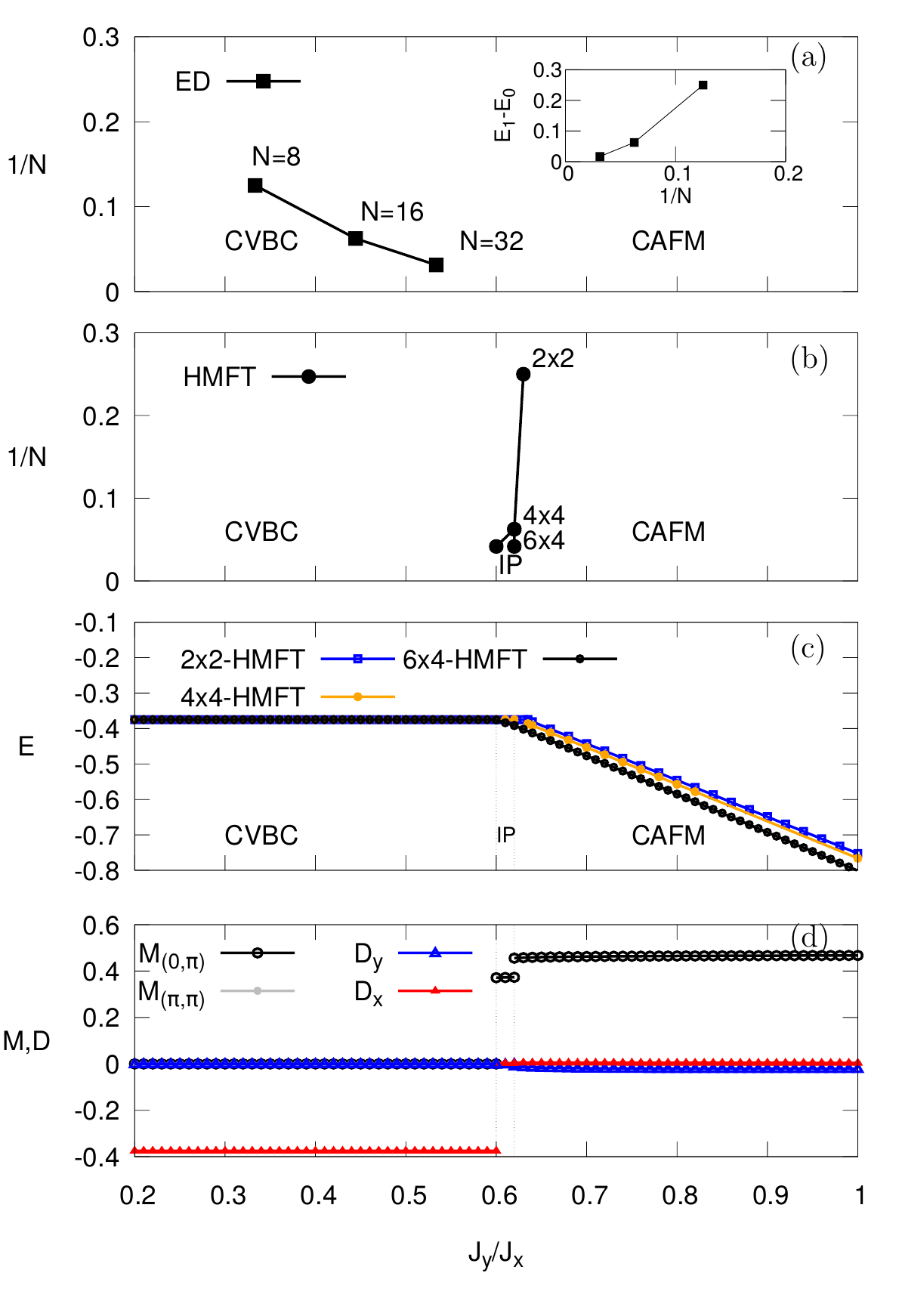}
\caption{\label{fig:DEMcol}Scaling of the GS phase diagram  of the CVBC parent Hamiltonian (\ref{eq:Hcol}) as computed with (a) ED and (b) HMFT. Inset (a): Singlet-triplet gap computed with ED at $J_y/J_x=1$.  (c) Energy per site (units of $J_x$) computed with HMFT for different cluster sizes of square shape. (d) Magnetization $M$ as computed with 6$\times$4-HMFT.}
\end{figure}

Upon tuning the anisotropy in the antiferromagnetic regime, we find for both Hamiltonians that the VBC phase extends beyond the exact GS regime and transits to a CAFM phase characterized by a vanishing singlet-triplet gap (ED) consistent with a strong columnar magnetization $M_{(0,\pi)}\sim 0.4$ (HMFT) through a region where various IPs appear when using large cluster sizes (HMFT) (see Figs.~\ref{fig:DEMstag} and \ref{fig:DEMcol}). 
In Fig.\ref{fig:DEMstag} we show the scaling of the GS phase diagram of the SVBC parent Hamiltonian (\ref{eq:Hstag}) as computed with ED ($N=8,16,32,36$) and HMFT ($N=4,8,16,24$), together with HMFT results on energies and order parameters ($M$ and $\mathcal{D}$). 
Both the ED (Fig.\ref{fig:DEMstag}a) and HMFT (Fig.\ref{fig:DEMstag}b) scalings of the phase diagram show that the SVBC boundary tends towards the near the isotropic regime.
Within HMFT and small cluster sizes ($N=4,8$), we find a first order SVBC-CAFM transition. 
For bigger cluster sizes ($N=16,24$), the direct SVBC-CAFM transition is replaced by a window of IPs: one within 16-HMFT and two (IP1 and IP2) within 24-HMFT. 
Interestingly, the IPs energies tend to smoothen the intersection of the SVBC and CAFM linear energies found with the smaller clusters (Fig.\ref{fig:DEMstag}c). 
The IP1 is characterized by suppression of dimer order, $\mathcal{D}_x\simeq -0.16$ (Fig.\ref{fig:DEMstag}d) while the IP2 is characterized by suppression of columnar magnetization. 
Note that these IPs are numerically converged only when restricting the variational space to cluster configurations with $\sum_j\langle S_j^z\rangle=0$, and that the transition IP2-CAFM takes place at a tiny region not fully converged even with this cutoff, both features signaling the presence of strong quantum fluctuations in this window.

In Fig. \ref{fig:DEMcol} we show the scaling of the GS phase diagram of the CVBC parent Hamiltonian (\ref{eq:Hcol}) as computed with ED ($N=8,16,32$) and HMFT ($N=4,16,24$) obtaining a similar picture.
Interestingly, in this case the HMFT phase diagram is quite stable from 2$\times$2 to 4$\times$4, and a single IP with slightly suppressed CAFM order and negligible dimerization appears in a small window when computing with the largest cluster, 6$\times$4-HMFT.
Same as for the SVBC parent Hamiltonian, this IP is only converged by applying the previously described cutoff to the variational space.
%

\textit{Summary and conclusion}.--- We have presented a systematic method to construct translation-invariant SU(2) symmetric VBC parent Hamiltonians based on a canonical mapping identifying a VBC state with the vacuum of a new set of dimer fermions (DFs).
This mapping guides our search for translation-invariant interactions that added to an anisotropic AF Heisenberg Hamiltonian render the VBC to be an exact eigenstate, and the exact GS when the related local Hamiltonian is semi-positive definite.
We have constructed the parent Hamiltonians of a staggered- and a columnar-VBC (SVBC and CVBC) on the square lattice, both containing frustrating four-spin AF interactions.
ED and HMFT-Gutzwiller calculations show a common phase diagram where the VBC extends over the exact regime and transits to a columnar AF (CAFM).
A window of strongly fluctuating intermediate phases (IPs) appear at the VBC-CAFM transition when computing with large clusters in HMFT.
Their subtle numerical convergence signal that this window might be governed by unusual characteristic length scales exceeding the cluster sizes used, and excludes a direct VBC-CAFM first order transition.
Greater system sizes, which lie beyond the limit of current computer capabilities, are needed to unveil the ultimate nature of this region.
Hints about the eventual proximity to an RVB state may be obtained by computing entanglement entropies, something which is beyond the scope of this work.
The method here presented can be applied to other 2D and 3D lattices of current experimental interest where quantum paramagnets are observed, such as the kagome~\cite{Norman2016}, for which sign-problem free QMC~\cite{Wildeboer2012,Wildeboer2016} and tensor network~\cite{Iqbal2014} calculations on RVB states have been already developed. 
Particularly promising in understanding quantum paramagnets would be the construction of Hamiltonians that may interpolate from a VBC to an RVB state.

\acknowledgments
We gratefully acknowledge J. Riera for providing us with exact diagonalization codes.
We acknowledge useful discussions with Adriana Foussats, Jorge Dukelsky, Gerardo Ortiz, Sumiran Pujari, Hui Shao, and Anders Sandvik. 
DH acknowledges computing time at JURECA, J\"ulich supercomputing center.
CG acknowledges support from CONICET-PIPI0389. 
AG acknowledges support from CONICET-PIP0375. 
We devote this paper to the memory of Prof. Dr. Alejandro Muramatsu, who died before this work was completed. 
Alejandro initiated this work developping the main ideas and formalism. 
We will miss him, his physics and friendship.

\bibliography{VBCMajoranas}

\pagebreak
\widetext
\begin{center}
\textbf{\large Supplemental Material}
\end{center}
\setcounter{equation}{0}
\setcounter{figure}{0}
\setcounter{table}{0}
\setcounter{page}{1}
\makeatletter
\renewcommand{\theequation}{S\arabic{equation}}
\renewcommand{\thefigure}{S\arabic{figure}}
\renewcommand{\bibnumfmt}[1]{[S#1]}
\renewcommand{\citenumfont}[1]{S#1}

\section{The dimer fermion mapping}
%
The vacuum and doubly occupied DF states, and the fully and singly occupied DF states, comprise two equivalent copies of the dimer Hilbert space. We can classify them by applying the total spin, $\mathbf{S}_r^2=(\mathbf{S}_{r,1}+\mathbf{S}_{r,2})^2$, and third component, $S^z_r=(S^z_{r,1}+S^z_{r,2})$, written in terms of DFs to each of the eight fermionic states of a dimer. 

The total spin of a dimer is,
\begin{equation}
\mathbf{S}_r^2=2\mathbf{S}_{r,1}\mathbf{S}_{r,2}+\mathbf{S}_{r,1}^2+\mathbf{S}_{r,2}^2,\label{eq:S_total}
\end{equation}
The first term of the right-hand side is indeed twice the intra-dimer Heisenberg (HB) interaction, $D_r=\mathbf{S}_{r,1}\mathbf{S}_{r,2}$, which can be written in terms of DFs by directly applying the mapping (2)-(3),
\begin{equation}
\mathbf{S}_{r,1}\mathbf{S}_{r,2}=
\frac{1}{4}\sum_a \sum_{b,c} \sum_{b',c'} \varepsilon^{abc}\varepsilon^{ab'c'}\left(T^{bc}_r+\Delta^{bc}_r\right)
\left(T^{bc}_r-\Delta^{bc}_r\right).
\end{equation}
Taking into account the properties of the Levi-Civita tensor, $\sum_a\varepsilon^{abc}\varepsilon^{ab'c'}=
\delta^{bb'}\delta^{cc'}
-\delta^{bc'}\delta^{cb'}$, that $T^{bc}_r\Delta^{bc}_r=0=\Delta^{bc}_rT^{bc}_r$, and using the antisymmetry properties of the $T$ (4) and $\Delta$ (5) DF operators,
\begin{equation}
\mathbf{S}_{r,1}\mathbf{S}_{r,2}
=\frac{1}{2}\sum_{b\neq c}\left( T^{bc}_rT^{bc}_r-\Delta^{bc}_r\Delta^{bc}_r \right).
\end{equation}
Substituting the number conserving ($T$) and non-conserving ($\Delta$) operators in terms of DFs and making use of the anticommutation relation of DFs,
\begin{equation}
\mathbf{S}_{r,1}\mathbf{S}_{r,2}
=-\frac{1}{8}\sum_{b\neq c} \left[
1-2 \left(f^{b\dag}_{r} f^b_{r} +f^{c\dag}_{r} f^c_{r}\right)\right]-\frac{1}{2}\sum_{b\neq c} 
f^{b\dag}_{r} f^b_{r} f^{c\dag}_{r} f^c_{r}.
\end{equation}
Finally, taking into account the three fermionic species,
\begin{equation}
\mathbf{S}_{r,1}\mathbf{S}_{r,2}=-\frac{3}{4} + \frac{1}{2} n_r \left(3-n_r \right).
\end{equation}

Following similar steps, one can show that $\mathbf{S}_{r,1}^2=3/4$ and $\mathbf{S}_{r,2}^2=3/4$ are satisfied when written in terms of DFs. Therefore, Eq. (\ref{eq:S_total}) has the final form,
\begin{equation}
\mathbf{S}_r^2=n_r(3-n_r),\label{eq:dimer_Tspin}
\end{equation}
where $n_r=\sum_a f_r^{a\dag}f^{a}_r$. 
The vacuum, singly, doubly, and fully occupied DF states are eigenstates of (\ref{eq:dimer_Tspin}) with eigenvalues 0,2,2, and 0, respectively.

The third component of the dimer spin, $S^z_r =(S^z_{r,1}+S^z_{r,2})$, can be equivalently obtained by directly applying the DF mapping (2)-(3),
\begin{equation}
S^z_r =2T^{xy}_r.
\end{equation}
Writing explicitly the number-conserving operator $T$ in terms of DFs,
\begin{equation}
S^z_r =-{\rm i}(f^{x\dag}_rf^{y}_r-f^{y\dag}_rf^{x}_r).\label{eq:Sz_DF}
\end{equation}
Applying the third component to the vacuum and doubly occupied states
\begin{eqnarray}
S^z_r\ket{000}&=&0,\\
S^z_r\ket{011}&=&-{\rm i}\ket{101},\label{eq:t_1}\\
S^z_r\ket{101}&=&{\rm i}\ket{011},\label{eq:t_2}\\
S^z_r\ket{110}&=&0
\end{eqnarray}
where we have used $\ket{n^x,n^y,n^z}$ with $n^a=f^{a\dag}f^{a}$ $(a=x,y,z)$ to refer to the dimer states in terms of the DF occupation basis. Collecting the eigenvalues of the total spin (\ref{eq:dimer_Tspin}) and its third component we can unambiguously identify the vacuum with the singlet, $\ket{s}=(1/\sqrt{2})(\ket{\uparrow\downarrow}-\ket{\downarrow\uparrow})$, 
\begin{equation}
\ket{000}=\ket{s}, 
\end{equation}
and $\ket{110}$ with the non-magnetic triplet, $\ket{t_0}=(1/\sqrt{2})(\ket{\uparrow\downarrow}+\ket{\downarrow\uparrow})$,
\begin{equation}
\ket{110}=e^{\rm{i}\phi}\ket{t_0},\label{eq:110}
\end{equation}
where $\phi$ is a phase to be determined below.

Diagonalizing the subspace defined by (\ref{eq:t_1}) and (\ref{eq:t_2}) we can identify the other two triplet states, $\ket{t_+}=\ket{\uparrow\uparrow}$ and $\ket{t_-}=\ket{\downarrow\downarrow}$, by
\begin{eqnarray}
\ket{t_+}&=& \frac{e^{\rm{i}\varphi_+}}{\sqrt{2}}(-\rm{i}\ket{101}+\ket{011}),\\
\ket{t_-}&=&\frac{e^{\rm{i}\varphi_-}}{\sqrt{2}}(\rm{i}\ket{101}+\ket{011}).
\end{eqnarray}
where $\varphi_+$ and $\varphi_-$ are phases that will be determined in the following by applying the non-conserving number operator $\Delta$ in terms of spin operators to the vacuum --i.e. the singlet state. 
In particular, by noticing that $\ket{110}=2{\rm i}\Delta^{xy}_r\ket{000}$ and that $(S^z_{r,1}-S^z_{r,2})=2\Delta^{xy}_r$ we have that 
\begin{equation}
\ket{110}={\rm i}(S^z_{r,1}-S^z_{r,2})\ket{s}=i\ket{t_0}.
\end{equation}
Comparing with (\ref{eq:110}), then $\phi=\pi/2$. Following analogous arguments and recalling that $S^x=(S^++S^-)/2$ and $S^y=(S^+-S^-)/2{\rm i}$ where $S^\pm$ stand for the ladder operators of the SU(2) group,
\begin{eqnarray}
\ket{101}&=&\frac{1}{\sqrt{2}}(\ket{t_+}+\ket{t_-}),\\
\ket{011}&=&-\frac{\rm i}{\sqrt{2}}(\ket{t_+}-\ket{t_-}).
\end{eqnarray}

Finally, inverting the relations we have
\begin{eqnarray}
\ket{s}&=&\ket{000},\\
\ket{t_0}&=&-{\rm i}\ket{110},\\
\ket{t_+}&=&\frac{1}{\sqrt{2}}(\ket{101}+{\rm i}\ket{011}),\\
\ket{t_-}&=&\frac{1}{\sqrt{2}}(\ket{101}-{\rm i}\ket{011}).
\end{eqnarray}

Equivalently, the fully and singly occupied states comprise a second equivalent copy of the singlet and triplet states.

\section{Local Hamiltonians of the SVBC and CVBC parent Hamiltonians}
%
The parent Hamiltonian of a VBC can be written as a translation-invariant sum of local Hamiltonians shifted by the DF vacuum eigenvalue, $H=\varepsilon+\sum_{j}H_j$, where $\varepsilon=-3J_x\mathcal{N}/8$.
In particular, the local Hamiltonian corresponding to the SVBC parent Hamiltonian (12) is
\begin{equation}
H_j= \frac{3}{8}J_x+
\frac{J_x}{2}B_{j,j+\hat{x}}
+\frac{J_y}{2}B_{j,j+\hat{y}} 
+J_2(B_{j,j+\hat{x}+\hat{y}} +B_{j+\hat{x},j+\hat{y}}) 
+QB_{j,j+\hat{x}}B_{j+\hat{y},j+\hat{x}+\hat{y}},
\end{equation}
that has a zero eigenvalue for $-4/3\ge J_y/J_x\ge 2/3$ when the second-NN HB and four-spin interaction strengths are fixed to $J_2=J_y/2$ and $Q=2J_x/3$ respectively, as indicated in the main text.

The local Hamiltonian of the CVBC parent Hamiltonian (13) is
\begin{eqnarray}
H_j&=&\frac{3}{8}J_x+
\frac{J_x}{4}B_{j,j+\hat{x}}
+\frac{J_y}{3}B_{j,j+\hat{y}} 
+\frac{J_2}{2}(B_{j,j+\hat{x}+\hat{y}} +B_{j+\hat{x},j+\hat{y}}) 
\notag\\
&&
+J_4 (B_{j,j+2\hat{x}+\hat{y}} +B_{j+\hat{y},j+2\hat{x}})
+\widetilde{Q}_x (B_{j,j+\hat{x}}B_{j+\hat{x}+\hat{y},j+2\hat{x}+\hat{y}} 
+ B_{j+\hat{x},j+2\hat{x}}B_{j+\hat{y},j+\hat{x}+\hat{y}}
),
\end{eqnarray}
that has zero eigenvalue for $J_y=0$ when the interactions are fixed to $J_2=J_y$, $J_4=J_y/2$, and $\widetilde{Q}_x=J_x/3$, as indicated in the main text.

\section{Hierarchical mean field theory and Exact diagonalization}
%
In the HMFT-Gutzwiller approach, the ansatz wave function is taken to be an uncorrelated product of cluster states,
\begin{equation}
\ket{\Psi}=\prod_R \ket{\Phi}_R,
\end{equation}
where $R$ represents the position of each cluster in the superlattice.
In this work, we use an homogeneus ansatz, i.e. all clusters to be equivalent, and thus we may drop the superlattice index, i.e. $\ket{\Phi}_R= \ket{\Phi}$.
The determination of the cluster state, $\ket{\Phi}=\sum_\alpha U_\alpha \ket{\alpha}$ with $\alpha$ representing cluster spin configurations in the $S^z$ basis, is obtained through variational optimization of the energy, which reduces to perform ED on a single cluster with open boundary conditions (OBC) an a set of self-consistently defined mean-fields acting on the boundaries.
It contains therefore unbiased information about competing orders with characteristic correlation lengths lying within the cluster dimensions.
The self-consistently defined mean-fields allow for the explicit breakdown of symmetries and the concomitant stabilization of long-range order.
Consequently, ED on the cluster is performed without implementing symmetries of the Hamiltonian, and therefore the cluster sizes used are smaller than in the standard ED with periodic boundary condition (PBC) procedures, due to memory limitations. 
For a given cluster shape, different mean-field configurations are used to seed the optimization procedure and the solution is obtained through comparing converged energies at each point of the phase diagram.
Increasing the cluster size allows to assess the validity of the result obtained with the inmediate smaller cluster.

\begin{figure}[b]
\includegraphics[width=0.9\textwidth]{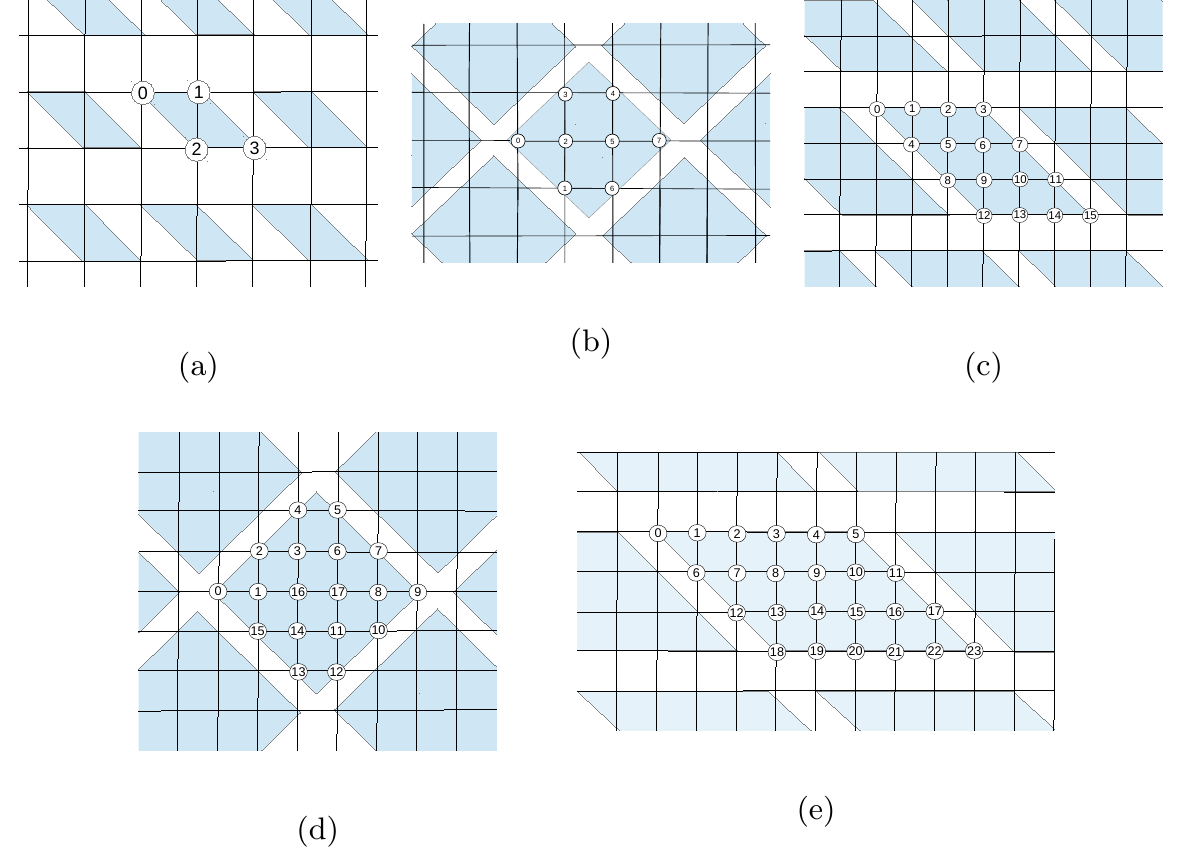}
\caption{\label{fig:hmft_clusters} Schematic representation of the cluster tiling resulting from the (a) 4- , (b) 8- , (c) 16- , (d) 18-, and (e) 24-HMFT Gutzwiller approach to the SVBC parent Hamiltonian. Clusters are schematically represented as blue shades. 
Labelled circles represent spins of the central cluster that is diagonalized self-consistently with the embedding mean-field bath resulting from the interaction with nearby clusters. 
All cases except for the 18-HMFT tiling (d) are commensurate with both the CAFM and SVBC phases. The 18-HMFT tiling of the square lattice is not commensurate with the CAFM order.} 
\end{figure}

In Figure S\ref{fig:hmft_clusters} we show the romboid and diamond shaped clusters used within HMFT-Gutzwiller to study the SVBC parent Hamiltonian ground state phase diagram.
The 18-HMFT cluster is not commensurate with the CAFM, although it is commensurate with the SVBC pattern.
Similarly, the clusters used within ED are commensurate with CAFM, AFM and VBC patterns.
In particular, diamond shaped $N=8,32$, and square shaped $N=16,36$ clusters are used to approach the SVBC parent Hamiltonian.
Same clusters are used to study the CVBC, except for the $N=36$: due to the shape of the four-spin interactions in the CVBC parent Hamiltonian, the maximum size that can be attained within ED computations is $N=32$ in this case.
Greater sizes commensurate with the SVBC and CVBC states ($N=50$, diamond shape) lie beyond current memory capabilities.

\end{document}